\documentclass[fleqn,twoside]{article}
\usepackage{espcrc2}

\usepackage{graphicx}
\usepackage[figuresright]{rotating}

\def\ij{\langle ij\rangle}

\newcommand{\AmS}{{\protect\the\textfont2
  A\kern-.1667em\lower.5ex\hbox{M}\kern-.125emS}}

\title{Phase structure of $(2+1)d$ strongly coupled lattice gauge theories} 

\author{Costas G. Strouthos \address{Department of Physics, Box 90305, 
Duke University, Durham, North Carolina 27708, USA.}
}

\begin{document}

\begin{abstract}
\noindent
We study the chiral phase transition in $(2+1)d$ strongly coupled $U(N)$ lattice
gauge theories with staggered fermions. We show with high precision simulations
performed directly in the chiral limit   
that these models undergo a Berezinski-Kosterlitz-Thouless (BKT) transition.
We also show that this universality class is unaffected even in the large N limit. 

\vspace{1pc}
\end{abstract}

\maketitle

\section{Introduction}
The behavior of symmetries at finite temperature is one of the most outstanding problems
in field theory. In recent years we have witnessed revised interest in the chiral phase
transition in QCD. The problem of symmetry breaking and its restoration is 
intrinsically non-perturbative and therefore most of our knowledge about the phenomenon 
comes from lattice simulations. 
However, computing quantities in lattice QCD with massless
quarks is a notoriously difficult problem, because most known algorithms break down in the 
chiral limit. In addition, the simulations near $T_c$ must be performed on lattices with large spatial 
sizes in order to control large finite size effects due to the diverging correlation
length. 

A useful simplification of QCD occurs in the strong coupling limit, which retains much of 
the underlying physics of QCD except for large lattice artifacts. 
In this limit chiral symmetry breaking and its restoration at finite temperature have 
been studied using large $N$ and large $d$ expansions \cite{kawamoto}. However, since these 
approaches are based on 
mean field analysis they cannot help in determining the universality of phase transitions.

Interestingly, lattice QCD 
with one staggered fermion interacting with $U(N)$ gauge fields can be mapped into a monomer-dimer
system in the strong coupling limit \cite{rossi}. Here, we present numerical results for the finite temperature
critical behavior of the $(2+1)d$ model. Our data were generated with a recently developed very efficient 
cluster algorithm \cite{adams}, 
which allows us to perform precision calculations in the chiral limit. 
In agreement with expectations from universality and dimensional reduction,
we show convincingly \cite{shailesh} that the chiral phase trasition belongs to the   
BKT universality class \cite{kosterlitz}. 

The partition function of the model we study here is given by
\begin{equation}
Z = \sum_{[n,b]} 
\prod_{\ij} (z_{\ij})^{b_{\ij}}\frac{(N-b_{\ij})!}{b_{\ij}! N!} 
\prod_i \frac{N!}{n_i!}m^{n_i},
\label{pf}
\end{equation}
and is discussed in detail in \cite{rossi,adams}.
 Here $n_i=0,1,2,...,N$
refers to the number of monomers on the site $i$,  $b_{\ij}=0,1,2,...,N$
represents the number of dimers on the bond $\ij$, $m$ is the monomer weight,
$z_{\ij}=\eta_{ij}^2$ are the dimer weights. Note that while spatial dimers
carry a weight $1$, temporal dimers carry a weight $T$. The sum is over
all monomer-dimer configurations $[n,b]$ which are constrained such that
the sum of the number of monomers at each site and the dimers that touch
the site is always $N$ (the number of colors). 
In this work we choose $L_x=L_y=L$. One can study the thermodynamics of the model
by working on asymmetric lattices with
$L_t \ll L$ and allowing $T$ to vary continuously.

\section{Results}

\begin{table*}[]
\caption{Fits for $\chi_c$ and $\chi_w$ near $T_c$.}
\label{table:1}
\renewcommand{\tabcolsep}{1.5pc} 
\renewcommand{\arraystretch}{0.9} 
\begin{tabular}{|c|c|c|c|c|c|c|}
\hline
$T$  &   $\eta$ &      2r    & $\chi^2_1/$d.o.f & $1/(2c)$  &   $\chi^2_2/$d.o.f \\
\hline
1.00 &  0.222(5) &  0         & 0.2   &    0.2343(8)   &   183.2   \\ 

1.02 &  0.235(5) &  -0.3(2)  & 1.5   &   0.2411(5)  &   53.4   \\

1.04 &  0.251(5) &  -0.07(2)  &  0.4   &   0.2483(5)  &   2.8    \\ 

1.06 &  0.249(5) & -0.03(2)  &  0.5   &   0.2583(5)  &   33.8   \\

1.10 &   0.388(5) &  -0.12(2) &  3.6   &  0.2831(5)  &   770.0  \\

1.14 &   0.569(6) &  -1.24(3) &  480   &   ---       &   ---    \\
\hline
\end{tabular}
\end{table*}

In this section we present the results for the $T \neq 0$
critical behavior of the model. The observables used in this work 
are the chiral susceptibility $\chi_c$ and the winding number susceptibility 
$\chi_w$. The latter is proportional to the helicity modulus and describes the response
of the system to a perturbation that distorts the direction of the spontaneous
magnetization. The winding number susceptibility has been used successfully in 
to demonstrate BKT behavior in other models \cite{harada}.
We fixed $L_t=4$ and computed $\chi_c$  and $\chi_w$ as a function of $L$. 
If $T_c$ is the critical temperature, then the BKT theory predicts
\begin{equation}
\chi_c \propto \left\{\begin{array}{lc}
L^{2-\eta(T)} & T<T_c \cr
L^{1.75}\ [\log(L)]^{0.125} & T=T_c \cr
\mbox{constant} & T>T_c  \cr
\end{array}\right.
\end{equation}
and
\begin{equation}
\chi_W = \left\{\begin{array}{lc}
1/[2\eta(T)] + \alpha_1L^{-\alpha_2} & T<T_c \cr
[2 + 1/\log(L/L_0)]                  & T=T_c \cr
\alpha_3 \exp{(-\alpha_4L)}          & T>T_c \cr
\end{array}\right.
\end{equation}
in the large $L$ limit.
The critical exponent $\eta(T)$ is expected to change continuously 
with T, but remains in the range $0\leq \eta(T) < 0.25$. In order to 
confirm these predictions we computed $\chi_c$ and $\chi_w$ for lattices
ranging from $L=32$ to $L=750$ and for $N=1,...,32$.

\begin{figure}[htb]
\centering
\includegraphics[scale=0.56]{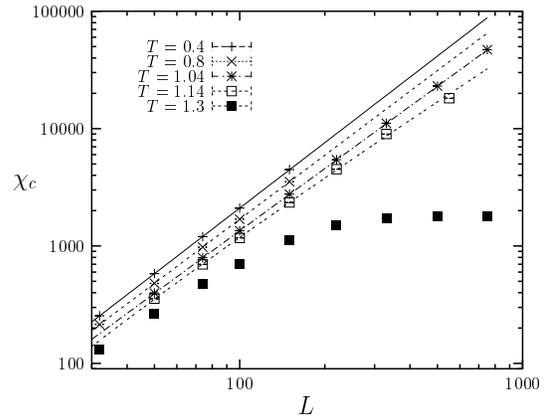}
\vspace{-1cm}
\caption{Plot of $\chi_c$ vs. $L$ for various $T$.}
\label{fig:all_Nf}
\end{figure}

\begin{figure}[b!]
\centering
\includegraphics[scale=0.56]{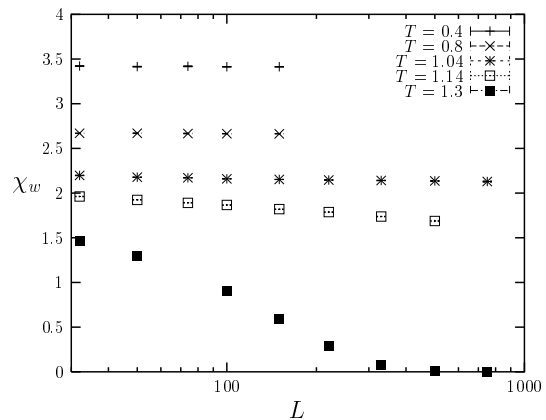}
\vspace{-1cm}
\caption{Plot of $\chi_w$ vs. $L$ for various $T$.}
\label{fig:all_Nf}
\end{figure}

Let us first discuss our results for $N=1$. 
In figures 1 and 2 we plot $\chi_c$ and $\chi_w$ as functions
of $L$ for different values of $T$. We find that $\chi_c$ fits well to the form 
$bL^{2-\eta}(\log(L))^{-2r}$
when $T \leq 1.06$. We also find that the logarithmic term is unimportant for 
$T \leq 1.0$, whereas at $T=1.04$ the value $2r=0.7(2)$ is close 
to the BKT prediction which is $0.125$. The values of $\eta$, 
$2r$ and the quality of the fits $\chi_1^2$/d.o.f are shown in table 1.
We also fit the data for $\chi_w$ to the form $(c+1/\log(L/L_0))$. The values of 
$1/(2c)$ are also shown in the table. 
Finally, using the fact that $\chi_w = (2+1/\log(L/L_0))$ is exactly valid at $T_c$ 
we fit the data for $\chi_w$ to this form for various values of $T$.
The values of $\chi_2^2$/d.o.f are shown in the last column of table 1. 
Based on where the minimum in $\chi_2^2$/d.o.f occurs we estimate $T_c=1.040(5)$. 
Our value of $\eta=0.250(5)$ at $T_c$ is in excellent agreement 
with the BKT prediction.

We also checked that $\chi_c$ and $\chi_w$ show similar evidence for a BKT 
transition at larger values of $N$. 
Using techniques similar to the ones we used for $N=1$ we 
computed $T_c$ for various values of $N$. We find
that $T_c = 0.708(6) N + 1.40(4) - 1.07(4)/N$ fits our results very
well for all values of $N$ with a $\chi^2$/d.o.f of $1.1$. The dependence 
of the coefficients of this polynomial on $L_t$ is still under investagation.
\begin{figure}[t!]
\centering
\includegraphics[scale=0.25]{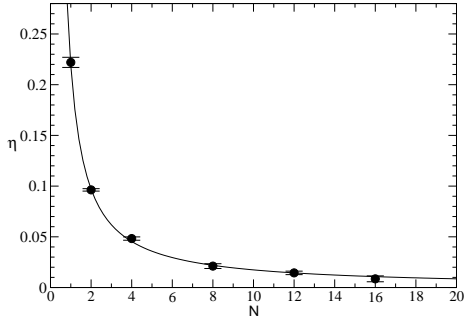}
\vspace{-1cm}
\caption{Plot of $\eta$ as a function of $N$ for $T=1.0$. 
The data fit very well to the form $\eta = 0.169(6)/N + 0.050(9)/N^2$ with 
$\chi^2$/d.o.f$=1.2$.}
\label{fig:all_Nf}
\end{figure}

\begin{figure}[htb]
\centering
\includegraphics[scale=0.56]{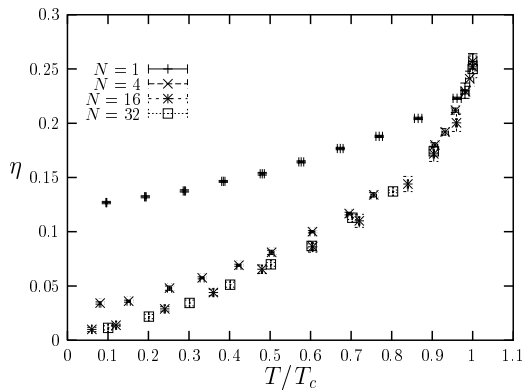}
\vspace{-1cm}
\caption{Plot of $\eta$ as a function of $T/T_c$ for various values of $N$.}
\label{fig:all_Nf}
\end{figure}
With regards to the $N$ dependence of our results we find two interesting 
obervations. Witten argued that when the symmetry is $U(1)$ the large $N$
analysis is still applicable since $\eta \sim 1/N$ at large $N$ \cite{witten}. Our results
for a fixed $T=1.0$, shown in figure 3, do agree with his conjecture. 
Interestingly, as $N$ becomes large and $T/T_c$ is held fixed instead of $T$, 
we find that $\eta \neq 0$ even in the large $N$ limit. Figure 4 shows that $\eta$
approaches an interesting function of $T/T_c$ as $N$ becomes large. Extending this 
observation to QCD, we think that the t'Hooft limit (large $N$ with $g^2N$ held fixed)
may be quite similar \cite{thooft}.

\section{Summary}
We presented high precision results from simulations of $(2+1)d$ $U(N)$ strongly coupled 
lattice gauge theories at $T\neq 0$
and we showed convincing evidence that the models undergo a BKT phase transition. 
In addition, we showed that this universality class is unaffected even in the large $N$ limit, implying 
that the mean field analysis often used in this limit breaks down in the critical region.

\section*{Acknowledgements}
\noindent
This work was done in collaboration with Shailesh Chandrasekharan and it was supported
in part by the NSF grant DMR-0103003 and DOE grant DE-FG-96ER40945.

\end{document}